\documentclass[aps,pra,twocolumn,groupedaddress,showpacs,showkeys]{revtex4}
\usepackage{amstext}
\usepackage{amsmath}
\usepackage{graphicx}
\usepackage{bm}

\begin{document}

\title{Entanglement conditions for two-mode states: Applications}

\author{Mark Hillery}
\affiliation{Department of Physics, Hunter College of CUNY, 695 Park Avenue,
New York, NY 10021}
\author{M. Suhail Zubairy}
\affiliation{Institute for Quantum Studies and Department of Physics,
Texas A\&M University, College Station, TX 77843}
\date{\today}

\begin{abstract}
We examine the implications of several recently derived conditions 
[Hillery and Zubairy, Phys.\ Rev.\ Lett.\ {\bf 96}, 050503 (2006)] for determining when a 
two-mode state is entangled.  We first find examples of non-Gaussian states that satisfy
these conditions.  We then apply the entanglement conditions 
to the study of several linear devices, the beam splitter, the parametric amplifier, and the
linear phase-insensitive amplifier.  For the first two, we find conditions on the input states that
guarantee that the output states are entangled.  For the linear amplifier, we determine in
the limit of high and no gain, when an entangled input leads to an entangled output.  Finally,
we show how application of two two-mode entanglement conditions to a three-mode state
can serve as a test of genuine three-mode entanglement.
\end{abstract}

\maketitle

\section{Introduction}
While quantum information theory was originally formulated in terms of qubits, higher-dimensional,
and, in particular, continuous-variable systems have shown considerable promise in applications
such as EPR correlation \cite{ou,silberhorn,bowen,howell} and quantum cryptography \cite{hillery1,ralph,reid}.   For a recent review see \cite{braunstein2}.
In view of the important role played by entanglement in
quantum information, this has led to an investigation of entanglement in continuous-variable
systems.  In particular, sufficiency conditions for multimode states to be entangled have been
formulated.  The first set of conditions that was found is both necessary and sufficient for entanglement in
Gaussian states and sufficient for entanglement in any two-mode state
\cite{simon,duan}. These conditions are expressed in terms of quantities that are at most 
quadratic in mode creation and annihilation operators.
They have been used, for example, to study entanglement in correlated emission laser systems \cite{xiong,tan}.  More recently, sufficient conditions for entanglement have
been found for a wider class of states \cite{hillery}-\cite{schukhin}.  Our aim here is to explore
some of the implications of these conditions.  We shall first discuss some non-Gaussian
states for which they can be used to detect entanglement.  We shall then go on to
use these conditions to study the behavior of entanglement in linear optical devices.  We shall
investigate what properties an input state to two particular optical
devices, beam splitters and parametric amplifiers (or degenerate four-wave mixers) needs to
have in order for the output state to be entangled.  We will also see how entanglement fares
when a state is amplified by a linear, phase-insensitive amplifier. Finally, we shall show how
two-mode entanglement conditions can be used to detect three-mode entanglement.

Nha and Kim have shown that the entanglement conditions developed in
\cite{hillery,agarwal} are useful in detecting entanglement in one
class of non-Gaussian states, $SU(2)$ minimum-uncertainty states 
\cite{nha}.  In particular, they demonstrated that the condition in
\cite{agarwal} was more effective for this purpose than a comparable
condition in \cite{hillery}.  They also discussed how the quantities
appearing in these conditions can be measured.

The entanglement produced by a beam splitter has been studied in considerable detail by
Kim, \emph{et al.} \cite{kim}.  They showed that a necessary condition for the output field
of a beam splitter to be entangled is that the input be nonclassical.  They then went on
to examine the entanglement produced by a number of nonclassical inputs, including squeezed
states and number states.  There is, however, more to be learned.  For example, it
was recently shown by Ivan, et al. that if the input state to a beam splitter is a product state of
the vacuum in one mode and a state with sub-Poissonian statistics in the other, then the output
state will be entangled \cite{ivan}.  They derived this result using the PPT (positive partial
transpose) condition \cite{peres,horodecki}.  We shall show that this result follows from one of the 
inequalities in \cite{hillery,schukhin}.  We shall then show how this result can be generalized by
applying a number of the other inequalities derived in these references.

By studying both the parametric and linear amplfiiers, we we gain information about the behavior
of entanglement under both phase-sensitive and phase-insensitive amplification.  This gives us
some insight into the production of ``bright'' entangled light.
That entanglement can be produced by a parametric amplifier is well-known, but the 
entanglement conditions in \cite{hillery} allow us to gain information about which input states
will produce it.  The linear amplifier does not produce entanglement, and, in fact, can destroy
it, and we will be able to find some conditions under which it can be preserved.

\section{Specific states}
Consider two modes, whose annihilation operators are $a$ and $b$.  The number operators for
each mode are $N_{a}=a^{\dagger}a$ and $N_{b}=b^{\dagger}b$.  We shall first concentrate on
the entanglement condition
\begin{equation}
\label{ent1}
|\langle ab^{\dagger}\rangle |^{2} > \langle N_{a}N_{b}\rangle  ,
\end{equation}
that is, if the two modes are in a state for which the above inequality is satisfied, the state is
entangled \cite{hillery}.  The quantities in this inequality can be measured in a relatively
straightforward way.  The quantity on the right-hand side can be measured by photon-counting
measurements.  The quantity on the left-hand side can be measured with the aid of a phase
shifter and a beam splitter.  Suppose the $b$ mode is first sent through a phase shifter that
performs the action $b\rightarrow e^{-i\phi }b$, and then both modes are sent into a beam 
splitter that sends $a\rightarrow (a+b)/\sqrt{2}$ and $b\rightarrow (b-a)/\sqrt{2}$.  We then
have that the output operators are
\begin{eqnarray}
a_{out} & = & \frac{1}{\sqrt{2}}(a + e^{-i\phi}b)   \nonumber \\
b_{out} & = & \frac{1}{\sqrt{2}}(-a + e^{-i\phi}b)  ,
\end{eqnarray}
and the expectation of difference of the numbers in the two modes at the output is given by
\begin{equation}
\langle(a_{out})^{\dagger}a_{out} - (b_{out})^{\dagger}b_{out} \rangle = e^{-i\phi}
\langle a^{\dagger}b\rangle + e^{i\phi}\langle a b^{\dagger}\rangle .
\end{equation}
By choosing $\phi =0$ we can measure the real part of $\langle a b^{\dagger}\rangle$ and by
choosing $\phi = -\pi /2$ we can measure the imaginary part.  These can then be combined to
yield $|\langle a b^{\dagger}\rangle |^{2}$.  It should be noted that higher-order field correlation
functions can also be measured, and explicit methods for doing so have been proposed by
Shchukin and Vogel \cite{vogel1}.

In order to gain a better understanding of the types of entangled states for which this condition
can be used to demonstrate entanglement, we shall study several examples. 
Note that if the two modes are in coherent states, that is the two-mode state is $|\alpha\rangle_{a}|\beta\rangle_{b}$, then we have that 
$|\langle ab^{\dagger}\rangle |^{2} = \langle N_{a}N_{b}\rangle $.  One way of possibly finiding
states that satisfy our entanglement condition, Eq.\ (\ref{ent1}), is to start with a product of
coherent states and perturbing this state in such a way as to produce entanglement.  We 
give two examples of this type.

The first example is that of a two-mode photon-added coherent state
\begin{equation}
|\psi\rangle_{ab} = \frac{1}{(|\alpha + \beta |^{2} +2)^{1/2}}(a^{\dagger}+b^{\dagger})
|\alpha\rangle_{a} |\beta\rangle_{b} ,
\end{equation}
which is not a Gaussian state.
Single-mode photon-added coherent states were first studied by Agarwal and Tara \cite{tara},
and they have recently been produced in the laboratory \cite{zavatta}.  For this state we find that
\begin{eqnarray}
\langle N_{a}N_{b}\rangle -|\langle ab^{\dagger}\rangle |^{2} & = & \frac{1}{|\alpha + \beta |^{2} +2}
[-4|\alpha |^{2}|\beta |^{2} 
\nonumber \\
& & -(\alpha^{\ast}\beta + \alpha \beta^{\ast})(|\alpha |^{2}+|\beta |^{2})
\nonumber \\
 & & -2(\alpha^{\ast}\beta + \alpha \beta^{\ast}) -1] .
\end{eqnarray}
If this quantity is negative, the state is entangled, and we can see that if we choose $\alpha$
and $\beta$ such that $(\alpha^{\ast}\beta + \alpha \beta^{\ast}) > 0$ this will indeed be the case.
Therefore, the entanglement condition in Eq.\ (\ref{ent1}) is capable of detecting entanglement
in some two-mode photon-added coherent states.

As a second example, let us consider a symmetric superposition of two two-mode coherent
states
\begin{equation}
|\psi\rangle_{ab} = \frac{1}{\sqrt{2}(1+|\langle\alpha |\beta\rangle |^{2})^{1/2}}(|\alpha\rangle_{a}
|\beta\rangle_{b} + |\beta\rangle_{a} |\alpha\rangle_{b}) ,
\end{equation}
which is, again, non-Gaussian.  For this state we find that
\begin{eqnarray}
\langle N_{a}N_{b}\rangle -|\langle ab^{\dagger}\rangle |^{2} & = & \frac{1}{4(1+x)^{2}}
 \{ -(\alpha^{\ast}\beta - \alpha \beta^{\ast})^{2}
\nonumber \\
& &  2x [ 4|\alpha\beta |^{2}-(\alpha\beta^{\ast}+\alpha^{\ast}\beta )(|\alpha |^{2}
 \nonumber  \\
& & +|\beta |^{2})] - x^{2}(|\alpha |^{2}-|\beta |^{2})^{2} \} ,
\end{eqnarray}
where $x=|\langle\alpha |\beta\rangle |^{2}$.
We see that if we choose $\alpha\beta^{\ast}$ to be real and positive, then this becomes
\begin{eqnarray}
\langle N_{a}N_{b}\rangle -|\langle ab^{\dagger}\rangle |^{2} & = & \frac{1}{4(1+x)^{2}}
[-2x|\alpha \beta | (|\alpha |-|\beta |)^{2} \nonumber \\
& & - x^{2}(|\alpha |^{2}-|\beta |^{2})^{2}] .
\end{eqnarray}
 The right-hand side is then negative,
and the condition in Eq.\ (\ref{ent1}) tells us that the state is entangled.

Next, we consider a higher-order entanglement condition from \cite{hillery} and apply it
to a superpostion of two-mode number states.  We have that a state is entangled if 
\begin{equation}
\label{ent2}
|\langle a^{m}(b^{\dagger})^{n}\rangle |^{2}  > \langle(a^{\dagger})^{m}a^{m}(b^{\dagger})^{n}b^{n}\rangle  .
\end{equation}
Now, consider the state,
\begin{equation}
|\psi\rangle = \frac{1}{\sqrt{2}}(|k_{1}\rangle_{a} |k_{2}\rangle_{b} + |k_{2}\rangle_{a}
|k_{1}\rangle_{b} ) ,
\end{equation}
which is a superposition of two states that are products of number states and is clearly 
entangled if $k_{1} \neq k_{2}$.  We shall assume
that $k_{1}>k_{2}$.  Choosing $m=n=k_{1}-k_{2}$, we find that
\begin{equation}
|\langle\psi |a^{(k_{1}-k_{2})}(b^{\dagger})^{(k_{1}-k_{2})}|\psi\rangle |^{2} =\frac{(k_{1}!)^{2}}
{4(k_{2}!)^{2}}  ,
\end{equation}
and
\begin{equation}
\langle\psi |(a^{\dagger})^{(k_{1}-k_{2})}a^{(k_{1}-k_{2})}(b^{\dagger})^{(k_{1}-k_{2})}
b^{(k_{1}-k_{2})} |\psi\rangle | = \frac{k_{1}!}{(2k_{2} -k_{1})!} ,
\end{equation}
if $2k_{2} \geq k_{1}$ and $0$ otherwise.  We see that if $2k_{2} < k_{1}$ the condition in
Eq.\ (\ref{ent2}) shows that the state is entangled, because the right-hand side is zero while
the left-hand side is positive.  If $2k_{2} \geq k_{1}$, then Eq.\ (\ref{ent2}) shows that the state
is entangled if 
\begin{equation}
k_{1}! (2k_{2}-k_{1})! >  4(k_{2}!)^{2}  .
\end{equation}
Note that because the entanglement conditions given in \cite{simon} and \cite{duan} contain
only quantities that are at most quadratic in mode creation and annihilation operators, they
will not be able to detect entanglement in this state if $k_{1}> k_{2}>2$.

Finally, let us show how we can increase the set of states for which entanglement can be
detected by modifying our entanglement condition.  The proof of the entanglement condition
in Eq.\ (\ref{ent1}) also goes through if we replace $a$ by $a-\langle a\rangle$ and $b$
$b-\langle b\rangle$, i.e.\ a state is entangled if
\begin{eqnarray}
\label{entmod1}
|\langle (a-\langle a\rangle )(b^{\dagger}-\langle b^{\dagger}\rangle)\rangle |^{2} > \nonumber \\
\langle  (a^{\dagger}-\langle a^{\dagger}\rangle )(a-\langle a\rangle )(b^{\dagger}-\langle b^{\dagger}\rangle)  (b-\langle b\rangle )\rangle  .
\end{eqnarray}
Now, suppose that we have a density matrix, $\rho$ for which $\langle a\rangle =\langle b\rangle
=0$, and which satisfies the condition in Eq.\ (\ref{ent1}).  Then the density matrix
\begin{equation}
\rho^{\prime}=D_{a}(\alpha )D_{b}(\beta )\rho D_{a}^{-1}(\alpha )D_{b}^{-1}(\beta ) ,
\end{equation}
where $D_{a}(\alpha )=\exp (\alpha a^{\dagger}-\alpha^{\ast}a)$ and $D_{b}(\beta ) =
\exp (\beta b^{\dagger} -\beta^{\ast}b)$ are mode displacement operators, satisfies our
modified entanglement condition.  As an example, consider the state 
\begin{equation}
|\Psi\rangle = \frac{1}{\sqrt{2}} (|0,1\rangle + |1,0\rangle ) , 
\end{equation}
which is a superposition of one photon in mode $a$ and one photon
in mode $b$.  It satisfies the entanglement condition in Eq.\ (\ref{ent1}), and $\langle a\rangle
=\langle b\rangle =0$. Therefore, the state 
\begin{equation}
|\Psi^{\prime}\rangle = D_{a}(\alpha )D_{b}(\beta )|\Psi\rangle  ,
\end{equation}
will satisfy the condition in Eq.\ (\ref{entmod1}), and 
can thereby be shown to be entangled.  This result
is not surprising, because the states differ only by local unitary transformations, and these will
not change the entanglement of a state.

\section{Optical devices}
We first discuss how a beam splitter acts on a two-mode field.  Suppose that the beam
splitter couples modes with annihilation operators $a$ and $b$.  The action of the beam splitter
can be described by a unitary operator, $U$, and the output operators, which we denote
by $a_{out}$ and $b_{out}$ are related to the input operators, which we denote simply by
$a$ and $b$, by $a_{out}=U^{\dagger}aU$ and $b_{out}=U^{\dagger}bU$.  The output operators
are related to the input operators by the relation \cite{mandel}
\begin{equation} 
\left( \begin{array} {c} a_{out} \\ b_{out} \end{array} \right) = \left( \begin{array}{cc} t & r \\
-r^{\ast} & t^{\ast} \end{array} \right) \left( \begin{array} {c} a \\ b  \end{array} \right) .
\end{equation}
The quantities $t$ and $r$ are the transmissivity and reflectivity of the beam splitter, respectively,
and they obey the relation $|r|^{2} +|t|^{2}=1$.

There are similar linear input-output relations for a degenerate parametric amplifier.  This is
also a device that couples two modes.  A pump mode, which is treated classically, provides
energy that allows the phase sensitive amplification of the two modes of interest, which are
called the signal and idler modes.  We denote their annihilation operators as $a$ and
$b$.  A similar interaction between two modes can be obtained by using four-wave mixing
with two strong, counter-propagating pump beams.  The output operators are related to the
input operators by \cite{mandel,sz}
\begin{eqnarray}
a_{out} & = & ca + s b^{\dagger} \nonumber   \\
b_{out} & = & cb + sa^{\dagger}  .
\end{eqnarray}
Here $c$ is real and positive, $s$ is complex, and they satisfy the relation $c^{2}-|s|^{2} = 1$. 
These numbers are related to the gain of the amplifier.

Now let us discuss our entanglement conditions.  A two-mode state is entangled if 
\begin{equation}
\label{cond1}
|\langle a^{m}(b^{\dagger})^{n}\rangle |^{2} > \langle (a^{\dagger})^{m}a^{m}(b^{\dagger})^{n}
b^{n}\rangle  ,
\end{equation}
or if 
\begin{equation}
\label{cond2}
|\langle a^{m}b^{n}\rangle |^{2} > \langle (a^{\dagger})^{m}a^{m}\rangle \langle (b^{\dagger})^{n}
b^{n}\rangle  ,
\end{equation}
for any integers $m,n \geq 1$ \cite{hillery}.  Here we shall be interested in the cases in which
$m=n$ and $m=1,2$.

Let us first consider the beam splitter.  If we assume that the input state is given by 
$|\Psi\rangle_{in}=|\psi\rangle_{a} |0\rangle_{b}$, that is an arbitrary state in the $a$ mode 
and the vacuum state in the $b$ mode, we find that 
\begin{eqnarray}
\langle ab^{\dagger}\rangle_{out} & = & -rt\langle a^{\dagger}a\rangle  \nonumber \\
\langle N_{a}N_{b}\rangle_{out} & = & |tr|^{2}( \langle N_{a}^{2}\rangle -\langle N_{a}\rangle ) ,
\end{eqnarray}
where $N_{a}=a^{\dagger}a$, $N_{b}=b^{\dagger}b$, and  expectation values with the subscript
 ``out'' are expectation values in the output state while those without a subscript are expectation
values in the state $|\Psi\rangle_{in}$.  If we now substitute these expressions into the condition 
in Eq.\ (\ref{cond1}) with $m=n=1$ we find that the output state is entangled if
\begin{equation}
\langle N_{a}\rangle > \langle N_{a}^{2}\rangle -\langle N_{a}\rangle^{2} =(\Delta N_{a})^{2} .
\end{equation}
This is the condition found by Ivan, \emph{et al.}, and it simply states that the output of the
beam splitter is in an entangled state if mode $a$ at the input has sub-Poissonian
photon statistics \cite{ivan}.  We can also use the same entanglement condition to see what
happens with a different input state.  Suppose the $b$ mode, rather than being in the vacuum
state is in a coherent state with amplitude $\beta$, i.e., $|\Psi\rangle_{in} = |\psi\rangle_{a} 
|\beta \rangle_{b}$.  We then have that
\begin{eqnarray}
|\langle ab^{\dagger}\rangle_{out}|^{2 }- \langle N_{a}N_{b}\rangle_{out} & = & |\beta |^{2} ( |r|^{4}
+ |t|^{4})(|\langle a\rangle |^{2} -\langle N_{a}\rangle ) \nonumber \\
 & & -(tr^{\ast})^{2} (\beta^{\ast})^{2} (\langle a\rangle^{2} - \langle a^{2}\rangle ) \nonumber \\
 & & -(t^{\ast}r)^{2}
\beta^{2} (\langle a^{\dagger}\rangle^{2} - \langle (a^{\dagger})^{2}\rangle ) 
\nonumber \\
& & + \mathcal{O}(\beta ) ,
\end{eqnarray}
where we have explicitly written down only the highest order terms in $\beta$, because we are
interested in the case in which the photon number in the $b$ mode is much larger than that in
the $a$ mode.    If we now express the complex quantities in terms of amplitudes and phases,
$t=|t|e^{i\theta_{t}}$, $r=|r|e^{i\theta_{r}}$, and $\beta = |\beta |e^{i\theta_{\beta}}$, and then set
$\phi =2(\theta_{t}-\theta_{r} -\theta_{\beta})$, and, in addition assume that $|t|=|r|=1/\sqrt{2}$, 
we have that 
\begin{eqnarray}
|\langle ab^{\dagger}\rangle_{out}|^{2 }- \langle N_{a}N_{b}\rangle_{out} & = & \frac{|\beta |^{2}} {4}
[ 2(|\langle a\rangle |^{2} -\langle N_{a}\rangle ) 
\nonumber \\
& & - e^{i\phi} (\langle a\rangle^{2} 
- \langle a^{2}\rangle ) \nonumber \\
& & - e^{-i\phi} (\langle a^{\dagger}\rangle^{2} - \langle (a^{\dagger})^{2}\rangle )] 
\nonumber \\
& & + \mathcal{O}(\beta ) .
\end{eqnarray}
We can choose a phase of $\beta$ to make this dominant term positive, and thereby
guaranteeing that the output state is entangled, if 
\begin{equation}
|\langle a^{2}\rangle -\langle a\rangle^{2} | > \langle N_{a}\rangle -|\langle a \rangle |^{2} .
\end{equation}
This is simply the condition that the $a$ mode be squeezed.  Therefore, if the input state consists of
a large-amplitude coherent state in one mode and a squeezed state with a much smaller photon
number in the other, we can adjust the phase of the coherent state to produce an entangled output
state.

We can also apply other entanglement conditions to the beam splitter output to find additional
kinds of input states that produce entangled output states.  We again consider input
states with the $a$ mode in an arbitrary state and the $b$ mode in the vacuum state.  If we
apply the condition in Eq.\ (\ref{cond1}) for $m=n=2$ to the output state resulting from such an
input, we find that 
\begin{eqnarray}
\langle a^{2}(b^{\dagger})^{2}\rangle_{out} & = & (rt)^{2} (\langle N_{a}^{2}\rangle 
-\langle N_{a} \rangle )  \nonumber \\
\langle (a^{\dagger})^{2}a^{2}(b^{\dagger})^{2}b^{2}\rangle_{out} & = & |tr|^{4}( \langle N_{a}^{2}
(N_{a}-1)^{2}\rangle 
\nonumber \\
& & -4\langle N_{a}(N_{a}-1)^{2}\rangle 
\nonumber \\ 
& & +2\langle N_{a}(N_{a}-1)\rangle )  ,
\end{eqnarray}
so that the output state is entangled if the input state satisfies the condtion
\begin{equation}
\langle N_{a}(N_{a}-1)\rangle^{2} > \langle N_{a}(N_{a}-1)(N_{a}-2)(N_{a}-3)\rangle .
\end{equation}
From this condition we can see that there are input states whose photon statistics are not 
sub-Poissonian, but which still lead to entangled states at the output.  A specific example 
is the state $(|0\rangle + |3\rangle )/\sqrt{2}$.
One can also apply the condition in Eq.\ (\ref{cond2}) with $m=n=1$ to the output state.  If we
again assume that the $b$ mode is initially in the vacuum state, we find that the output is 
entangled if the $a$-mode input state satisfies
\begin{equation}
|\langle a^{2}\rangle | > \langle N_{a}\rangle .
\end{equation}
This inequality is satisfied by a squeezed state that satisfies the additional condition
$\langle a \rangle =0$.

Next we consider the parametric amplifier.  We apply the condition in Eq.\ (\ref{cond2})
with $m=n=1$ and $m=n=2$.  Again we assume an input state of the form $|\Psi\rangle_{in}
=|\psi\rangle_{a} |0\rangle_{b}$.  The $m=n=1$ condition gives us that the output is entangled
if the input $a$ mode state satisfies (we assume here that $|s|>0$)
\begin{equation}
\label{paramp1}
c > |s|   ,
\end{equation}
which is always true.  Therefore, any input state for the $a$ mode will lead to an entangled
output state.  Higher-order conditions lead to the same conclusion.  The $m=n=2$
condition for entanglement gives us that the output is entangled if 
\begin{equation}
2\left( 1-  \frac{|s|^{2}}{c^{2}}\right) \langle N_{a}\rangle + \left(1-  \frac{|s|^{4}}{c^{4}} \right) >0 ,
\end{equation}
which is again satisfied for any $a$-mode input state.

Finally, let us see what the entanglement conditions in \cite{simon} and \cite{duan} tell us about
the inputs of beam splitters and parametric amplifiers.  In order to state these conditions, we first
define
\begin{eqnarray}
x_{a}=\frac{1}{\sqrt{2}} (a^{\dagger}+a) & \hspace{1cm}& x_{b}=\frac{1}{\sqrt{2}} (b^{\dagger}+b)
\nonumber \\
p_{a}=\frac{i}{\sqrt{2}} (a^{\dagger}-a) & \hspace{1cm}& p_{b}=\frac{i}{\sqrt{2}} (b^{\dagger}-b) .
\end{eqnarray}
Next, for any real $\xi$, define
\begin{eqnarray}
u & = & |\xi |x_{a}+\frac{1}{\xi}x_{b}  \nonumber \\
v & = & |\xi |p_{a}-\frac{1}{\xi}p_{b}  .
\end{eqnarray}
Finally, we can say that if a state satisfies the inequality
\begin{equation}
\label{ent3}
(\Delta u)^{2} + (\Delta v)^{2} < \xi^{2} + \frac{1}{\xi^{2}} ,
\end{equation}
then it is entangled.

If we assume that the input state of the beam splitter is of the form $|\psi\rangle_{a}|0\rangle_{b}$,
then we find that
\begin{eqnarray}
(\Delta u)_{out}^{2} + (\Delta v)_{out}^{2} & = & -\frac{2|\xi |}{\xi}[tr^{\ast}(\langle a^{2}\rangle
-\langle a\rangle^{2})
\nonumber \\
& & +t^{\ast}r(\langle (a^{\dagger})^{2}\rangle - \langle a^{\dagger}\rangle^{2})]
\nonumber \\
  & & +\left( |t\xi |^{2}+\frac{|r|^{2}}{\xi^{2}}\right) [2(\langle N_{a}\rangle 
\nonumber \\  
& &  -|\langle a\rangle |^{2})
+1]  \nonumber \\
 & & +\left( |r\xi |^{2}+\frac{|t|^{2}}{\xi^{2}}\right)  .
\end{eqnarray}
The output state will be entangled if
\begin{equation}
(\Delta u)_{out}^{2} + (\Delta v)_{out}^{2} - \xi^{2} - \frac{1}{\xi^{2}} < 0 .
\end{equation}
We now minimize the left-hand side with respect to $\xi$.  The result is that the output state
is entangled if
\begin{eqnarray}
& &\pm [tr^{\ast}(\langle a^{2}\rangle -\langle a\rangle^{2})
+t^{\ast}r(\langle (a^{\dagger})^{2}\rangle 
- \langle a^{\dagger}\rangle^{2})] 
\nonumber \\
& & +2|rt| (\langle N_{a}\rangle -|\langle a\rangle |^{2}) < 0,
\end{eqnarray}
where the plus or minus sign is chosen so as to minimized the left-hand side.  Suppose we
send the $a$ mode state through a phase shifter, which will send $a\rightarrow e^{-i\phi}a$,
before sending it into the beam splitter, and we can choose $\phi$ so as to minimize the
left-hand side of the above inequality.  We then find that the above entanglement condition
will be satisfied if
\begin{equation}
|\langle a^{2}\rangle -\langle a\rangle^{2}| > \langle N_{a}\rangle -|\langle a\rangle |^{2}  ,
\end{equation}
which is simply the condition that the input $a$ mode state be squeezed.

We can now apply the condition in Eq,\ (\ref{ent3}) to the parametric amplifier.  As usual, we 
assume that in the input state the $b$ mode is in the vacuum state.  The derivation is similar
to the one for the beam splitter, so we just give the result.  Defining 
\begin{equation}
\eta =2( \langle N_{a}\rangle -|\langle a\rangle |^{2} ) +1 ,
\end{equation}
we find that the output state is entangled if the input state satisfies
\begin{equation}
2|s|(\eta c^{2}+|s|^{2}-1)^{1/2} < c |s+s^{\ast}|(\eta +1)^{1/2} .
\end{equation}
This condition is more restrictive than the one derived from Eq.\ (\ref{cond2}) with $m=n=1$,
which showed that the output is entangled for any$a$-mode input state and any phase of
$s$.  If $s$ is real, then the above inequality is satisfied for any $a$-mode input state, but if
$s$ is imaginary, then it is never satisfied and gives us no information. 

\section{Linear amplifier}
A linear amplifier is a device that provides phase-insensitive amplification of an optical signal.
It will not create entanglement, but it is useful to see what happens to an input state that does
possess entanglement.  It also allows us to see how losses affect entanglement.

A linear amplifier for a single mode is described by the master equation \cite{mandel}
\begin{equation}
\frac{d\rho}{dt} = \mathcal{L}_{ga}(\rho ) + \mathcal{L}_{la}(\rho )  ,
\end{equation}
where $\rho$ is the density matrix of the mode, $\mathcal{L}_{ga}$ is the Liouville operator
describing the gain
\begin{equation}
\mathcal{L}_{ga}(\rho ) =\frac{A_{a}}{2}(2a^{\dagger}\rho a -aa^{\dagger}\rho -\rho aa^{\dagger}) ,
\end{equation}
and $\mathcal{L}_{la}$ is the Liouville operator describing the loss
\begin{equation}
\mathcal{L}_{la}(\rho ) = \frac{C_{a}}{2}(2a\rho a^{\dagger} -a^{\dagger}a\rho -\rho a^{\dagger}a) .
\end{equation}
For two modes, the master equation becomes
\begin{equation}
\frac{d\rho}{dt} = \mathcal{L}_{ga}(\rho ) + \mathcal{L}_{la}(\rho ) +  \mathcal{L}_{gb}(\rho ) 
+ \mathcal{L}_{lb}(\rho ) .
\end{equation}
From the master equation we can find equations of motion for expectation values of operators,
and the solution of these equations is straightforward.  We find that (expectation values at time
$t$ are denoted by a subscript $t$ and those at time $0$ are denoted by a subscript $0$)
\begin{eqnarray}
\langle ab^{\dagger}\rangle_{t} & = & e^{(A_{a}+A_{b}-C_{a}-C_{b})t/2} 
\langle ab^{\dagger}\rangle_{0}  \nonumber \\
\langle N_{a}N_{b}\rangle_{t} & = & e^{(A_{a}+A_{b}-C_{a}-C_{b})t} \langle N_{a}N_{b}\rangle_{0}
\nonumber \\
& & +\frac{A_{a}}{A_{a}-C_{a}} e^{(A_{b}-C_{b})t} (e^{(A_{a}-C_{a})t}-1) \langle N_{b}\rangle_{0}
\nonumber \\
 & & +\frac{A_{b}}{A_{b}-C_{b}} e^{(A_{a}-C_{a})t} (e^{(A_{b}-C_{b})t}-1)\langle N_{a}\rangle_{0}
\nonumber  \\
 & & + \frac{A_{a}}{A_{a}-C_{a}} \frac{A_{b}}{A_{b}-C_{b}}  (e^{(A_{a}-C_{a})t}-1)  \nonumber \\
 & & (e^{(A_{b}-C_{b})t}-1) .
\end{eqnarray}

Let us use these results to see how the entanglement condition in Eq.\ (\ref{ent1}) behaves upon
amplification, or attenuation, in two limiting cases.  First, let us set the gain terms to zero, i.e.\ 
$A_{a}=A_{b}=0$.  This allows us to see what happens when only losses are present.  We find
that 
\begin{eqnarray}
|\langle ab^{\dagger}\rangle_{t}|^{2}- \langle N_{a}N_{b}\rangle_{t}  & = & e^{-(C_{a}+C_{b})t}
(|\langle ab^{\dagger}\rangle_{0}|^{2}  \nonumber \\
& & - \langle N_{a}N_{b}\rangle_{0} ) .
\end{eqnarray}
From this we see that losses do not, in principle, affect our ability to detect entanglement by 
means of Eq.\ (\ref{ent1}); if the condition is satisfied initially it will be satisfied for any later
time.  It does, of course, become more and more difficult to detect the difference between the
two quantities appearing in this condition as time progresses.  Our second case is the high-gain
limit.  We shall assume that $A_{a}-C_{a}>0$ and $A_{b}-C_{b}>0$ and that $t$ is large, so that
we shall only keep terms proportional to $G_{ab}^{2}=\exp [(A_{a}+A_{b}-C_{a}-C_{b})t]$.  
We then find that 
\begin{eqnarray}
|\langle ab^{\dagger}\rangle_{t}|^{2}- \langle N_{a}N_{b}\rangle_{t} = G_{ab}^{2} [
(|\langle ab^{\dagger}\rangle_{0}|^{2}- \langle N_{a}N_{b}\rangle_{0} ) \nonumber \\
-\frac{A_{a}}{A_{a}-C_{a}}\langle N_{b}\rangle_{0} -\frac{A_{b}}{A_{b}-C_{b}}\langle N_{a}\rangle_{0} \nonumber \\
- \frac{A_{a}A_{b}}{(A_{a}-C_{a})(A_{b}-C_{b})}] .
\end{eqnarray}
If we now note that, for any state
\begin{equation}
|\langle ab^{\dagger}\rangle |^{2} \leq \langle N_{a}N_{b}\rangle + \langle N_{a}\rangle ,
\end{equation}
we see that the right-hand side of the above equation is always less than or equal to zero,
so that in the high-gain limit the condition in Eq.\ (\ref{ent1}) is no longer able to detect whether
there is entanglement in the output state.

The fact that our condition can no longer detect entanglement in the high-gain regime does not
necessarily mean that there is no entanglement there.  Using different arguments, however,
we can show that if the gain is too high any initial entanglement in the input state will be
absent at the output.  In order to do so we make use of a results due to Hong, Friberg,
and Mandel, which shows that  when the gain of a linear amplifier is too large, its output 
will be classical \cite{hong}.  If, for
a single mode, the P-representation of the input state to a linear amplifier is $P_{0}(\alpha )$,
then at time $t$ the P-representation is given by
\begin{equation}
P(\alpha ,t) = \int d^{2}\alpha^{\prime} P_{0}(\alpha^{\prime})\frac{1}{\pi m(t)}
e^{-|\alpha -G(t)\alpha^{\prime}|^{2}/m(t)} ,
\end{equation}
where
\begin{eqnarray}
G(t) & = & e^{(A-C)t/2} \nonumber   \\
m(t) & = & \frac{A}{A-C}(G(t)^{2} -1)  .
\end{eqnarray} 
They showed that if $G(t)^{2}\geq A/C$, then the output state is classical, i.e.\ $P(\alpha ,t)$ has
the properties of a probability distribution.  These results are easily extended to two modes.
In that case, if the input state has a P-representation given by $P_{0}(\alpha ,\beta )$, then the
P-representation at time $t$ is given by
\begin{eqnarray}
P(\alpha ,\beta ,t) & = & \int d^{2}\alpha^{\prime} \int d^{2}\beta^{\prime}P_{0}(\alpha^{\prime},
\beta^{\prime})\frac{1}{\pi^{2} m_{a}(t) m_{b}(t)}  \nonumber \\
& & e^{-|\alpha -G_{a}(t)\alpha^{\prime}|^{2}/m_{a}(t)} \nonumber  \\
& & e^{-|\beta -G_{b}(t)\beta^{\prime}|^{2}/m_{b}(t)}  ,
\end{eqnarray}
where $G_{a}(t)$ and $m_{a}(t)$ are given by the above expressions with $A$ and $C$
replaced by $A_{a}$ and $C_{a}$, and $G_{b}(t)$ and $m_{b}(t)$ are given by the above
expressions with $A$ and $C$ replaced by $A_{b}$ and $C_{b}$.  The P-representation of
the output state, $P(\alpha ,\beta ,t)$, will be classical if $G_{a}(t)^{2}\geq A_{a}/C_{a}$ and
$G_{b}(t)^{2}\geq A_{b}/C_{b}$.  This also means the output state will be separable.  We 
have that 
\begin{equation}
\rho_{ab}(t) = \int d^{2}\alpha \int d^{2}\beta P(\alpha , \beta , t) |\alpha\rangle_{a}\langle \alpha | 
\otimes |\beta \rangle_{b}\langle \beta |  ,
\end{equation}
where $|\alpha\rangle_{a}$ is a coherent state in the $a$ mode, and $|\beta\rangle_{b}$ is a coherent state in the $b$ mode.  From the above equation, it is clear
that if $P(\alpha , \beta , t)$ is a probability distribution, then $\rho_{ab}$ is separable.  Therefore,
we can conclude that if the single-mode gains are sufficiently large, the output state of the
amplifier will separable no matter how entangled the input was.  Linear, phase-insensitive
amplification with sufficiently high gain destroys entanglement.

\section{Three-mode entanglement}
We now briefly want to examine entanglement in a three-mode system.  We shall denote the
modes, and their respective annihilation operators, by $a$, $b$, and $c$. Entanglement conditions
for three-mode Gaussian states were formulated by Giedke, \emph{et al}. \cite{giedke}.   Conditions
for determining whether a general three-mode state is completely separable were give in 
\cite{hillery} and very recently conditions for multimode entanglement  have been studied by
Shchukin and Vogel \cite{vogel2}.  

In studying three-mode states, we are often interested in which subsystems are responsible for
the entanglement.  If the state is entangled, it may be the case that only two of the modes are
entangled, while the third is not entangled with either of these two modes.  For example, if the
density matrix is of the form
\begin{equation}
\label{form1}
\rho_{abc} = \sum_{j}p_{j}\rho_{aj}\otimes \rho_{bcj} ,
\end{equation}
where $0\leq p_{j}\leq 1$ and $\sum_{j}p_{j}=1$, then mode $a$ will not be entangled with either
mode $b$ or mode $c$, but modes $b$ and $c$ can be entangled leading to the overall 
entanglement of the state.  If a three-mode density matrix cannot be expressed in the above 
form, or in either of the two forms
\begin{eqnarray} 
\label{form2}
\rho_{abc} & = & \sum_{j}p_{j}\rho_{bj}\otimes \rho_{acj} \nonumber \\
\rho_{abc} & = & \sum_{j}p_{j}\rho_{cj}\otimes \rho_{abj} ,
\end{eqnarray}
then we say that it is genuinely entangled.  It has been shown how to produce genuinely 
entangled multimode states by van Loock and Braunstein \cite{braunstein}.

We now want to give some simple conditions for determining whether a three-mode state is
genuinely entangled, and to give an example of such a state that is not Gaussian and
whose entanglement can be demonstrated by these conditions.  Suppose that the 
three mode density matrix, $\rho_{abc}$ is of the form given in Eq.\ (\ref{form1}) or of the form
of the first line in Eq.\ (\ref{form2}).  Then it is the case that $\rho_{ab}={\rm Tr}_{c}(\rho_{abc})$ 
is separable, and the results of \cite{hillery} imply that it must satisfy
\begin{equation}
|\langle ab^{\dagger}\rangle |^{2} \leq \langle N_{a}N_{b}\rangle .
\end{equation}
Therefore, if $\rho_{abc}$ satisfies the condtion 
\begin{equation}
\label{gen1}
|\langle ab^{\dagger}\rangle |^{2} > \langle N_{a}N_{b}\rangle 
\end{equation}
it cannot be of either of these two forms.  Similarly, if it satisfies the condition
\begin{equation}
\label{gen2}
|\langle bc^{\dagger}\rangle |^{2} > \langle N_{b}N_{c}\rangle ,
\end{equation}
it cannot be of the form given in the second line of Eq.\ (\ref{form2}).  If it satisfies both of these
conditions, it must be genuinely three-mode entangled.

A simple example of a state that does satisfy these conditions is a three-mode, single-photon
W state
\begin{equation}
|\Psi\rangle = \frac{1}{\sqrt{3}}(|0,0,1\rangle + |0,1,0\rangle + |1,0,0\rangle ) .
\end{equation}
This state is a superposition of states in which one mode has one photon and the other two
modes are in the vacuum state.  For this state we find that $\langle N_{a}N_{b}\rangle = 
\langle N_{b}N_{c}\rangle = 0$, and $\langle ab^{\dagger}\rangle = \langle bc^{\dagger}\rangle
=1/3$.  Therefore, both of the above conditions are satisfied, and the state is genuinely
three-mode entangled.

By replacing the one-photon state in the above example with a coherent state, we can find a
family of states that is genuinely three-mode entangled.  That is we consider the state
\begin{eqnarray}
|\Psi (\alpha )\rangle  & = & \eta (|0\rangle_{a}|0\rangle_{b}|\alpha\rangle_{c}+ |0\rangle_{a} 
|\alpha \rangle_{b}|0\rangle_{c} \nonumber \\
& & + |\alpha \rangle_{a}|0\rangle_{b}|0\rangle_{c}) ,
\end{eqnarray}
where $|\alpha\rangle$ is a coherent state and 
\begin{equation}
\eta = \frac{1}{[3(1+2e^{-|\alpha |^{2}})]^{1/2}} .
\end{equation}
For this state we again have that $\langle N_{a}N_{b}\rangle = \langle N_{b}N_{c}\rangle = 0$, 
but now $\langle ab^{\dagger}\rangle = \langle bc^{\dagger}\rangle = |\eta \alpha |^{2}
 \exp (-|\alpha |^{2}|)$.  Therefore, we see that for all nonzero values of $\alpha$ the state 
$|\Psi (\alpha )\rangle$ exhibits genuine three-mode entanglement, though this entanglement
is easiest to detect for $|\alpha |\sim 1$, because that is when the difference between the two
sides of the inequalities, Eqs.\ (\ref{gen1}) and (\ref{gen2}), is greatest.

\section{Conclusion}
We have discussed a number of applications of the entanglement conditions derived in Ref
 \cite{hillery}.  We have given examples of non-Gaussian states whose entanglement can
be detected by these conditions.  We have also used them to study the entanglement produced,
or destroyed, by linear optical devices, in particular beam splitters, parametric amplifiers (with
a classical pump) and linear amplifiers.  Finally, we showed how these conditions could be
simply extended so that they can be used to detect genuine three-mode entanglement.  The
quantities in these entanglement conditions, at least the simplest ones, are relatively simple to
measure, and it should be possible to use them to detect entanglement in the laboratory.


\begin{thebibliography}{99}
\bibitem{ou} Z. Y. Ou, S. F. Pereira, H. J. Kimble, and K. C. Peng, Phys. Rev. lett. {\bf 68}, 3663 (1992).
\bibitem{silberhorn} Ch. Silberhorn, P. K. Lam, O. Wei$\beta$, F. K\"{o}nig, N. Korolkova1,
and G. Leuchs1 Phys. Rev. Lett. {\bf 86}, 4267 (2001).
\bibitem{bowen} W. P. Bowen, R. Schnabel, and P. K. Lam, Phys. Rev. Lett. {\bf 90}, 043601 (2003).
\bibitem{howell} J. C. Howell, R. S. Bennink, S. J. Bentley, and R. W. Boyd, Phys. Rev. Lett. {\bf 92}, 210403 (2004).
\bibitem{hillery1} M. Hillery, Phys. Rev. A {\bf 61}, 022309 (2000).
\bibitem{ralph} T. C. Ralph, Phys. Rev. A {\bf 61}, 010303 (2000).
\bibitem{reid} M. D. Reid, Phys. Rev. A {\bf 62}, 062308 (2000).
\bibitem{braunstein2} S.\ L.\ Braunstein and P.\ van Loock, Rev.\ Mod.\ Phys.\ {\bf 77}, 513 (2005).
\bibitem{simon} R.\ Simon, Phys.\ Rev.\ Lett.\ {\bf 84},2726 (2004).
\bibitem{duan} L.\ -M\ Duan, G.\ Giedke, J.\ I.\ Cirac, and P.\ Zoller, Phys.\ Rev.\ Lett.\ {\bf 84},
2722 (2000).
\bibitem{xiong} H. Xiong, M. O. Scully, and M. S. Zubairy, Phys. Rev. Lett. {\bf 94}, 023601 (2005)
\bibitem{tan} H.-T. Tan, S.-Y. Zhu, and M. S. Zubairy, Phys. Rev. A {\bf 72}, 022305 (2005).
\bibitem{hillery} M.\ Hillery and M.\ S.\ Zubairy, Phys.\ Rev.\ Lett.\ {\bf 96}, 050503 (2006).
\bibitem{agarwal} G.\ S.\ Agarwal and A.\ Biswas, New J.\ Phys.\ {\bf 7}, 211 (2005).
\bibitem{schukhin}E.\ Schukhin and W.\ Vogel, Phys.\ Rev.\ Lett.\ {\bf 95}, 230502 (2005).
\bibitem{nha} Hyunchul Nha and Jaewan Kim, quant-ph/0512180.
\bibitem{kim} M.\ S.\ Kim, W.\ Son, V.\ Bu\v{z}ek, and P.\ L.\ Knight,
Phys.\ Rev.\ A {\bf 65}, 032323 (2002).
\bibitem{ivan} J.\ S.\ Ivan, N.\ Mukunda, and R.\ Simon, quant-ph/0603255.
\bibitem{peres} A. Peres, Phys.\ Rev.\ Lett.\ {\bf 77}, 1413 (1996).
\bibitem{horodecki} M. Horodecki, P.\ Horodecki, and R.\ Horodecki, Phys.\ Lett.\ A  {\bf 223},
1 (1996).
\bibitem{vogel1} E.\ Schukhin and W.\ Vogel, Phys.\ Rev.\ A {\bf 72}, 043808 (2005).
\bibitem{tara} G.\ S.\ Agarwarl and K.\ Tara, Phys.\ Rev.\ A {\bf 43}, 492 (1991).
\bibitem{zavatta} A.\ Zavatta, S.\ Viciani, and M.\ Bellini, Science {\bf 306}, 660 (2004);
 A.\ Zavatta, S.\ Viciani, and M.\ Bellini, Phys.\ Rev.\ A {\bf 72}, 023820 (2005).
\bibitem{mandel} L.\ Mandel and E.\ Wolf, \emph{Optical Coherence and Quantum Optics} 
(Cambridge University Press, Cambridge, 1995).
\bibitem{sz} M. O. Scully and M. S. Zubairy, {\it Quantum Optics}, (Cambridge Press, London 1997).
\bibitem{hong} C.\ K.\ Hong, S.\ Friberg, and L.\ Mandel, J.\ Opt.\ Soc.\ Am.\ B, {\bf 2}, 494 (1985).
\bibitem{giedke} G.\ Giedke, B.\ Kraus, M.\ Lewenstein, and J.\ I.\ Cirac, Phys.\ Rev.\ A {\bf 64},
052303 (2001).
\bibitem{vogel2} E.\ Shchukin and W. Vogel, quant-ph/0605154.
\bibitem{braunstein} P.\ van Loock and S.\ L.\ Braunstein, Phys.\ Rev.\ Lett.\ {\bf 84}, 3482 (2000).

\end{thebibliography}
\end{document}